\documentclass[twocolumn]{aastex631}

\usepackage{graphicx}
\usepackage{amsmath}
\usepackage{mathtools}
\usepackage{amssymb}
\usepackage{hyperref}
\usepackage{float}
\newcommand{\code}[1]{\texttt{#1}}

\usepackage{graphicx}

\shortauthors{Arias et al.}

\graphicspath{{./}{figures/}}
\usepackage{graphicx} 
\usepackage{xfrac}

\begin{document}

\title{Andromeda XXXV: The Faintest Dwarf Satellite of the Andromeda Galaxy}

\correspondingauthor{Jose Marco Arias}
\email{joarias@umich.edu}

\author[0009-0002-9085-5928]{Jose Marco Arias}
\affiliation{Department of Astronomy, University of Michigan, 1085 S. University Ave., Ann Arbor, MI 48109-1107, USA}

\author[0000-0002-5564-9873]{Eric F. Bell}
\affiliation{Department of Astronomy, University of Michigan, 1085 S. University Ave., Ann Arbor, MI 48109-1107, USA}

\author[0000-0003-2294-4187]{Katya Gozman}
\affiliation{Department of Astronomy, University of Michigan, 1085 S. University Ave., Ann Arbor, MI 48109-1107, USA}

\author[0000-0002-2502-0070]{In Sung Jang}
\affiliation{Department of Astronomy and Astrophysics, University of Chicago, Chicago, IL 60637, USA}

\author[0000-0003-1677-0213]{Saxon Stockton}
\affiliation{Department of Physics, Utah Valley University, 800 W. University Parkway, Orem, UT 84058, USA}

\author[0000-0001-9852-9954]{Oleg Y.\ Gnedin}
\affiliation{Department of Astronomy, University of Michigan, 1085 S. University Ave., Ann Arbor, MI 48109-1107, USA}

\author[0000-0001-9269-8167]{Richard D'Souza}
\affiliation{Vatican Observatory, Specola Vaticana, V-00120, Vatican City State}

\author[0000-0003-2325-9616]{Antonela Monachesi}
\affiliation{Departamento de Astronom\'{i}a, Universidad de La Serena, Avda. R\'{a}ul Bitr\'{a}n 1305, La Serena, Chile}

\author[0000-0001-6380-010X]{Jeremy Bailin}
\affiliation{Department of Physics and Astronomy, University of Alabama, Box 870324, Tuscaloosa, AL 35487-0324, USA}

\author[0000-0002-1793-3689]{David Nidever}
\affiliation{Department of Physics, Montana State University, P.O. Box 173840, Bozeman, MT 59717-3840, USA}

\author[0000-0001-6982-4081]{Roelof S. de Jong}
\affiliation{Leibniz-Institut f\"{u}r Astrophysik Potsdam (AIP), An der Sternwarte 16, 14482 Potsdam, Germany}

\begin{abstract}
We present the discovery of Andromeda XXXV, the faintest Andromeda satellite galaxy discovered to date, identified as an overdensity of stars in the Pan-Andromeda Archaeological Survey and confirmed via Hubble Space Telescope imaging. Located at a heliocentric distance of $927^{+76}_{-63}$\,kpc and $158^{+57}_{-45}$\,kpc from Andromeda, Andromeda XXXV is an extended ($r_h = 53\,^{+13}_{-11}$ pc), elliptical ($\epsilon = 0.4\, \pm 0.2$), metal-poor ($[\text{Fe}/\text{H}]\sim-1.9$) system, and the least luminous ($M_V=-5.2 \pm 0.3$) of Andromeda's dwarf satellites discovered so far. Andromeda XXXV's properties are consistent with the known population of dwarf galaxies around the Local Group, bearing close structural resemblance to the Canes Venatici II and Hydra II Milky Way (MW) dwarf satellite galaxies. Its stellar population, characterized by a red horizontal branch or a red clump feature, mirrors that of other Andromeda satellite galaxies in showing evidence for a spread in age and metallicity, with no signs of younger stellar generations. This age-metallicity spread is not observed in MW satellites of comparable stellar mass, and highlights the persistent differences between the satellite systems of Andromeda and the MW, extending even into the ultrafaint regime. 
\end{abstract}
\keywords{Dwarf galaxies; Andromeda Galaxy}

\section{Introduction} \label{sec:intro}
\begin{figure*}[htb!]
\centering
\includegraphics[width=0.98\textwidth]{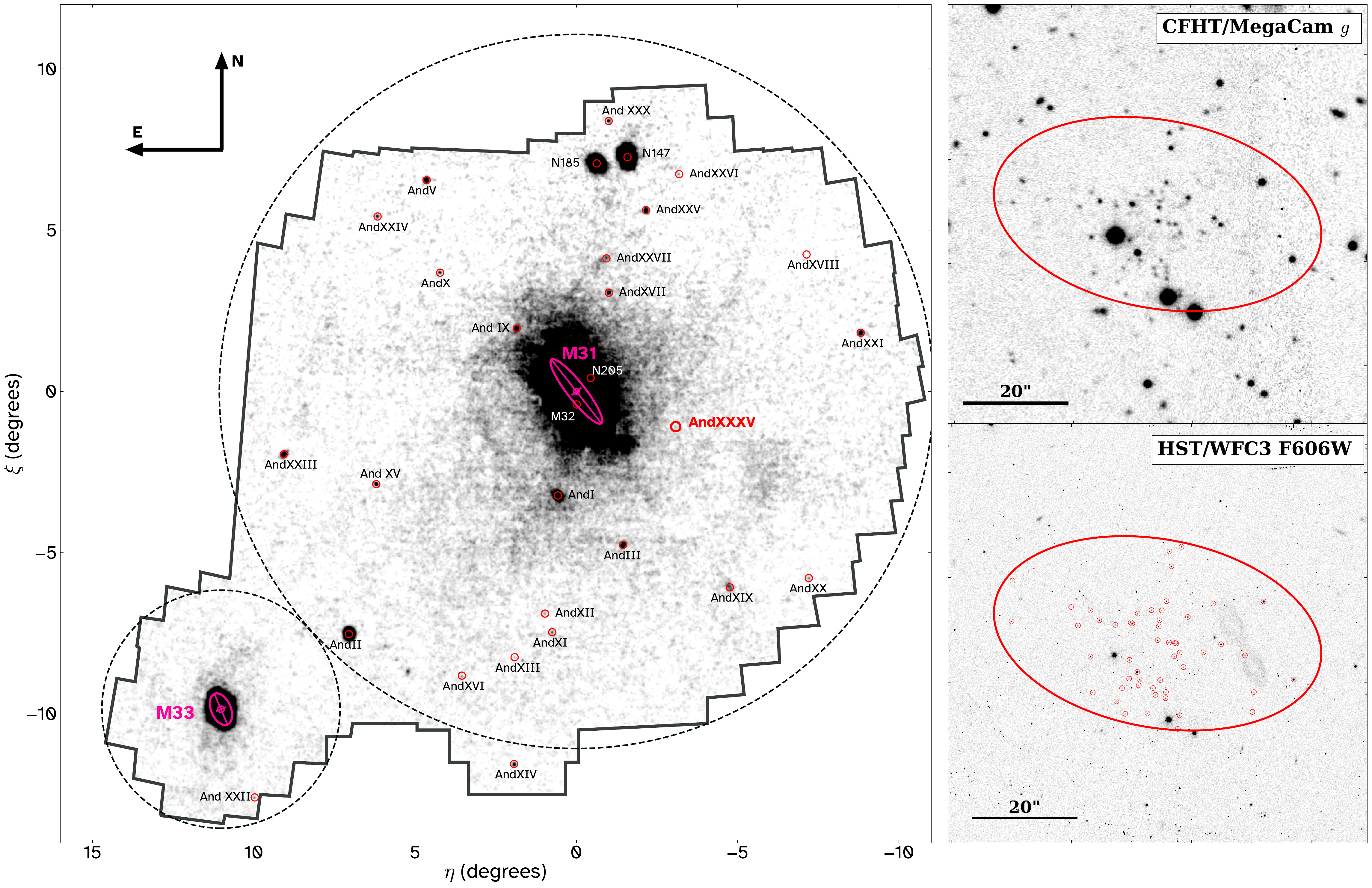}
\caption{Left: surface density map of candidate red giant branch stars with iron abundances between $-2.5 \lesssim [\text{Fe}/\text{H}] \lesssim -2$ at the distance of M31. This density map was created using point sources selected from the PAndAS `Combined stellar catalog' \citep{McConnachie2018}. The extent of the PAndAS footprint is represented by the thick black solid line. The coordinates on the tangent plane projection are $(\xi, \eta)$ centered on M31. The image has been smoothed using a top-hat filter with adaptive widths that depend on the local density in $1.2\arcmin \times 1.2\arcmin$ grid cells. Contaminants have been removed from this map by subtracting a supplementary map constructed from objects with colors and magnitudes indicative of foreground and background contamination sources. The dwarf galaxies within the PAndAS footprint are highlighted by red circles, with And XXXV distinguished by a thicker outline and bold red font. The dashed circles correspond to projected radii of 150 kpc from M31, and 50 kpc from Triangulum (M33). All major stellar substructures in the halo of M31 are visible. For reference, the names of these stellar substructures are provided in Figure 12 of \citet{McConnachie2018}. Upper right: PAndAS CFHT/MegaCam $g$-band discovery mosaic of And XXXV. Lower right: HST/WFC3 $\text{F606W}$ follow-up image of And XXXV. Red circles mark sources within And XXXV's best-fit ellipse that meet our photometric quality metrics for point sources. Both images are $80\arcsec \times 80\arcsec$. \label{fig:PAndAS_Dens_Map}}
\end{figure*} 

Ultrafaint dwarf galaxies (UFDs; $M_V>-7.7$; \citealt{Simon2019,Drlica-Wagner2020}) are critical probes of the properties of dark matter and the physical drivers of galaxy formation at very low mass scales \citep{Bullock2017}. In the roughly two decades since their first discovery (e.g., \citealt{belokurov2007}), tens of ultrafaint satellite galaxies of the Milky Way (MW) have been discovered \citep{Simon2019}. As individual systems, they are extremely dark matter dominated and most appear to have primarily early star formation that is impacted strongly by reionization \citep[e.g.,][]{Brown2014, Weisz_2014_SFHs_LG_DGs}. Many nearby UFDs appear to have been associated with the LMC, highlighting the importance of the infall of galaxy groups for delivering satellites \citep[e.g.,][]{DOnghia_Lake_2008_Samall_DGs_within_Large_DGs, Kallivayalil_2018_Missing_Sats_MagellClouds, Patel2020}. At the least luminous limits, the properties of ultrafaint galaxies and star clusters start to overlap, requiring careful study of the kinematics, metallicities, and sometimes abundance patterns of member stars to distinguish galaxies from star clusters---indeed, many systems remain ambiguous \citep[e.g.,][]{WillmanStrader2012,Simon2019,Fu_2023_NarrowBand_Met_13_UFD_Cands,Richstein2024}. Ultrafaint satellite galaxies are discovered as concentrations of resolved stars, and discovery space is limited entirely by the depth and point-spread function (PSF) of the discovery imaging surveys \citep[e.g.,][]{Koposov2008, Drlica-Wagner2020}. Only recently have sky surveys matured to allow the discovery of UFDs outside the MW in large sky surveys \citep[e.g.,][]{McQuinn2024_LeoK_LeoM}, or in smaller area deep sky surveys around nearby more massive galaxies \citep[e.g.,][]{Martin2013,Okamoto2019,Bell2022,Mutlu-Pakdil2022_NGC253}. 

\begin{figure*}[htb!]
\centering
\includegraphics[width=0.30\textwidth]{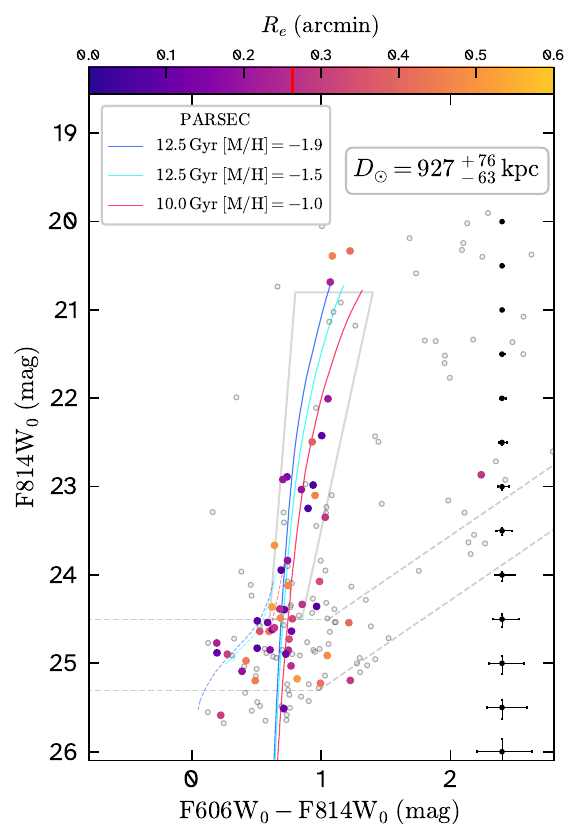}
\includegraphics[width=0.4\textwidth]{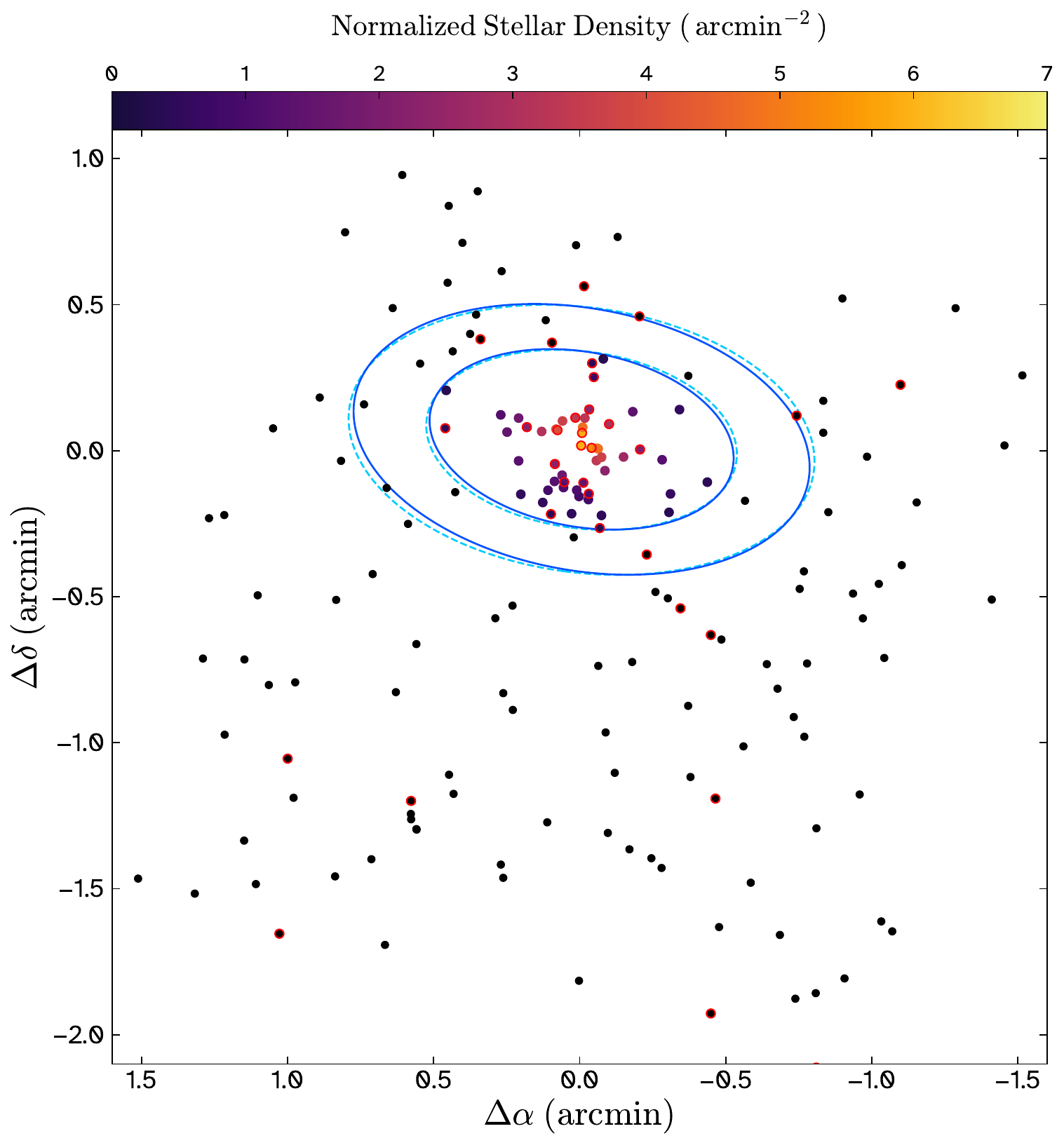}
\includegraphics[width=0.284\textwidth]{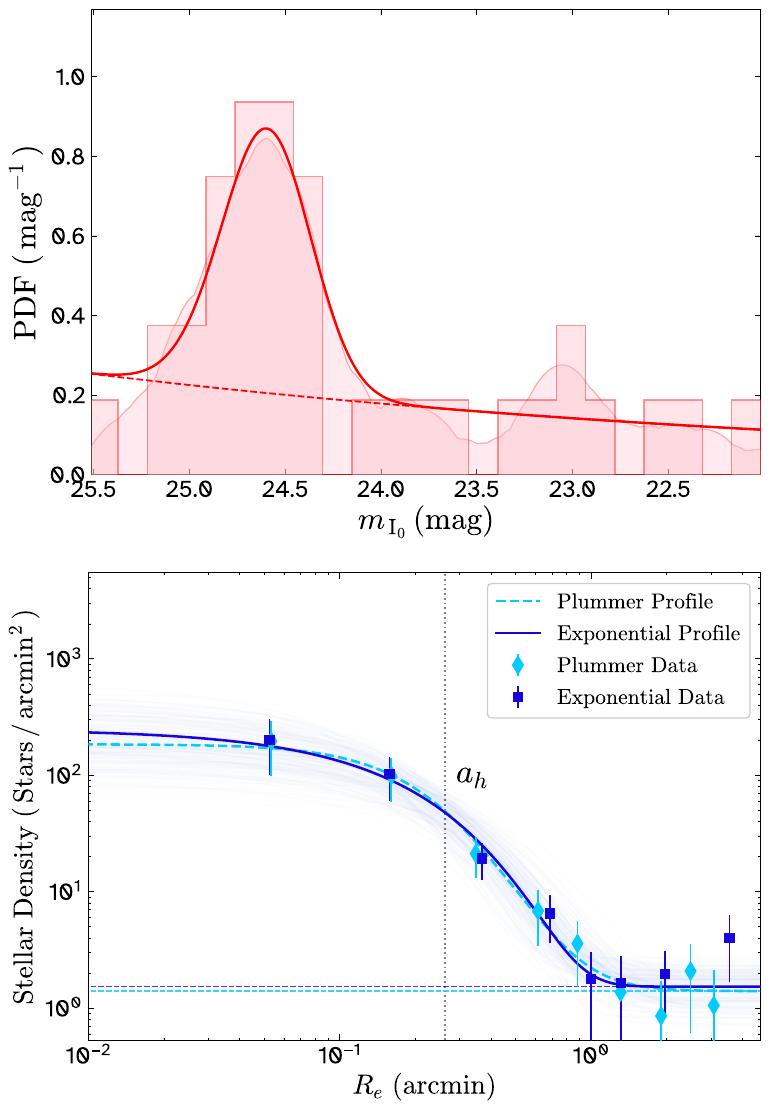}
\caption{Left: extinction-corrected CMD of stars within the HST/WFC3 field of view. Old (12.5\,Gyr) PARSEC isochrones \citep{2012_Parsec} with metallicities of $[\text{M}/\text{H}]=-1.9$ and $-1.5$\,dex are overlaid, along with a slightly younger (10\,Gyr) isochrone with $[\text{M}/\text{H}]=-1.0$\,dex. The corresponding horizontal branch tracks, shown as dashed lines in the same colors as their respective isochrones, are also included. All stars meeting our photometric quality metrics are shown as light gray dots, with those within $2 \times a_h$ colored by their elliptical radius from And XXXV's center. Candidate red horizontal branch stars are located around $\text{F606W}_0 -\text{F814W}_0 \sim 0.6$\,mag and $\text{F814W}_0 \sim 24.6$\,mag. The elliptical half-light radius $a_h$ of the best-fit exponential profile is marked by a red solid line in the colorbar. The CMD selection box used to derive the structural parameters of And XXXV is outlined by the light gray region, and the gray dashed lines represent the 80\% (50\%) completeness limits of the data. Middle: spatial distribution of all point sources within the HST/WFC3 field of view. Sources within $2\times a_h$ are colored based on the normalized stellar surface densities of the best-fit exponential model. The sources outlined in red denote the CMD-selected stars used in our MCMC analysis. The dark blue solid ellipses represent $2 \times a_h$ and $3 \times a_h$ from the system's centroid as defined by the exponential profile, and the dashed light blue ellipses indicate $2 \times r_p$ and $3 \times r_p$ for the Plummer profile. All structural parameters used to plot these ellipses are detailed in Table \ref{table:properties}. Top right: $I$-band luminosity function around the red horizontal branch feature of And XXXV. The observed distribution of stars is shown as a normalized histogram, while the solid red line represents the best-fit model (a Gaussian plus an exponential; see Section \ref{sec:distances}). An Epanechnikov kernel density estimate (light pink) is overplotted, highlighting the concentration of stars around the red horizontal brach feature. Bottom right: binned radial density profiles for And XXXV. The curves represent the best-fitting exponential (solid) and Plummer (dashed) density profiles. The light solid lines around the best-fit models represent 100 random draws from the posterior distributions of the free model parameters. The dashed horizontal lines denote the background densities (see Table \ref{table:properties}) for each model. The uncertainties are derived assuming Poisson statistics.\label{fig:Cand1_Diag_Dens_plot}}
\end{figure*}
\begin{figure*}[htb!]
\centering
\includegraphics[width=0.48\textwidth]{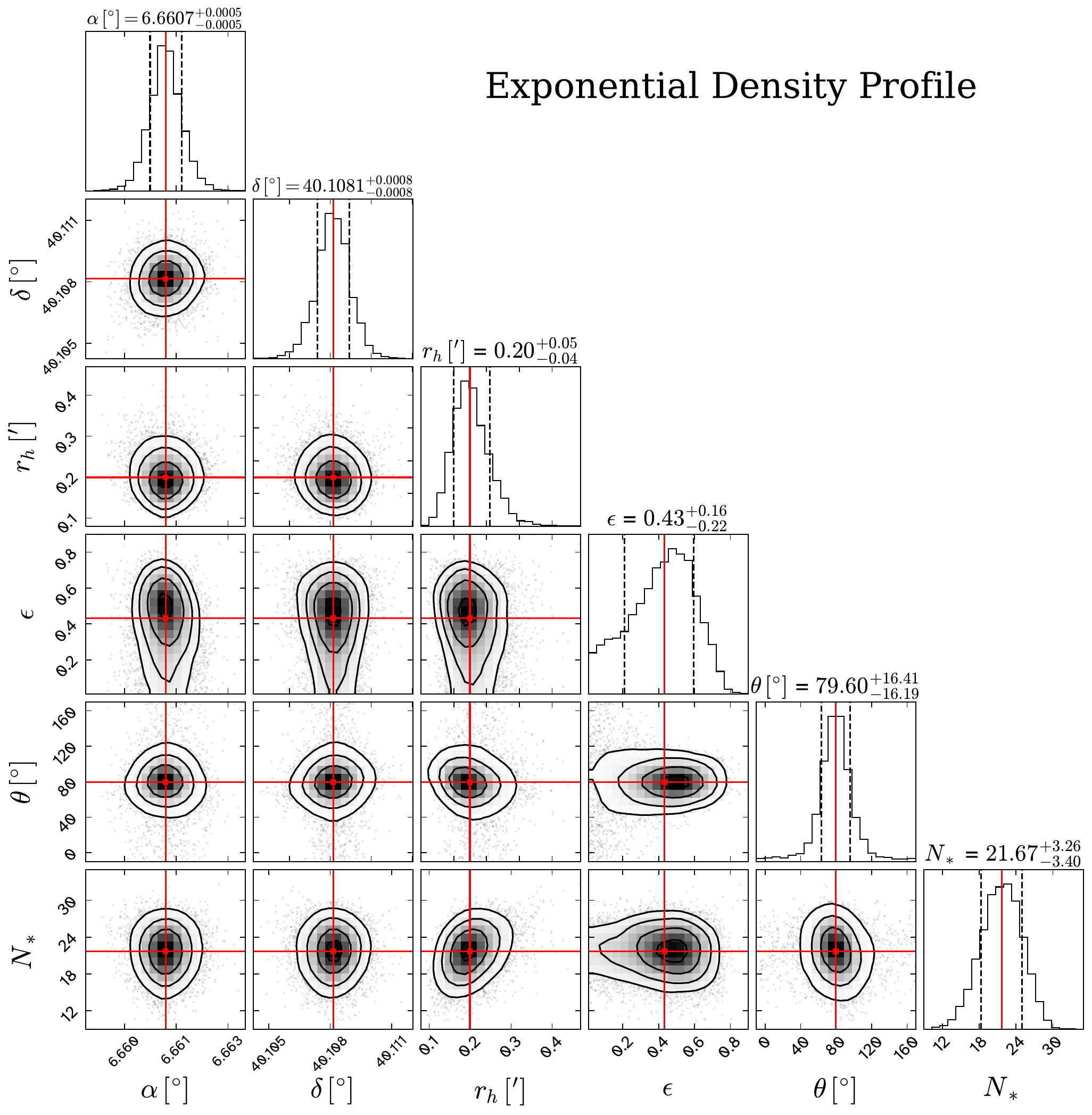}
\includegraphics[width=0.48\textwidth]{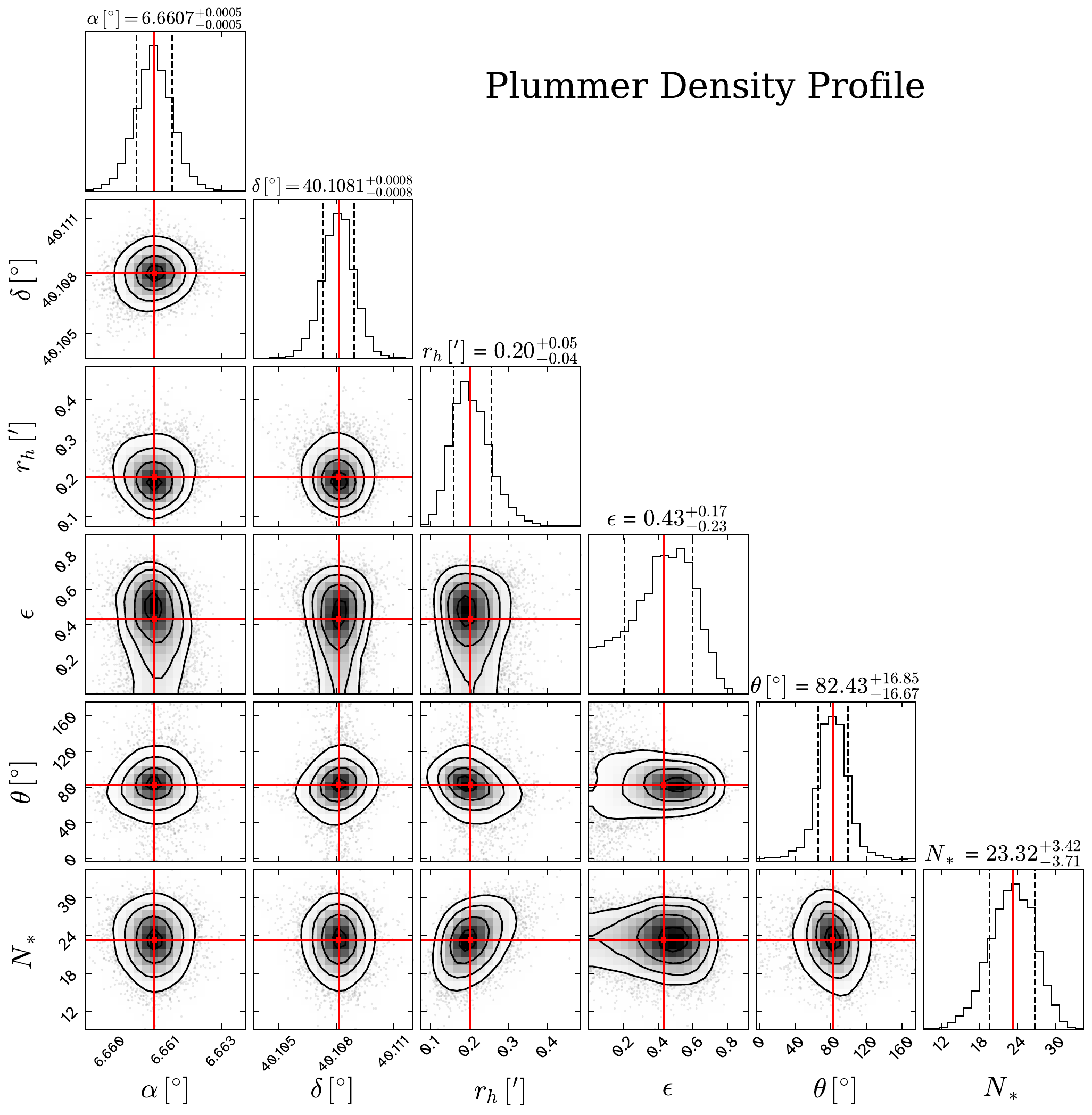}
\caption{Corner plots displaying the full posterior distributions of the structural parameters of the exponential (left) and Plummer (right) models for And XXXV. The azimuthally averaged half-light radius $r_h = a_h \sqrt{1-\epsilon}$ is displayed rather than the elliptical half-light radius $a_h$ for both density profiles. The centroid of And XXXV has already been converted to its R.A. (J2000) and decl. (J2000) coordinates in degrees. The posterior distributions for both corner plots are based on over 10,000 MCMC iterations after burn-in. The red solid lines and the dashed gray lines represent the medians and the 16th and 84th percentiles of each parameter, respectively. \label{fig:Cand1_corner_plots}}
\end{figure*}

These substantial developments in imaging capabilities of photometric surveys have now rendered the satellite system of Andromeda (M31) an especially fertile ground for the discovery and analysis of UFDs. The Pan-Andromeda Archaeological Survey (PAndAS; \citealt{McConnachie2018}) has significantly enriched our catalog of satellite galaxies around M31, uncovering nearly 20 dwarf galaxies \citep[e.g.,][]{Martin2006_AndXI-AndXIII,Martin2009_Pandas_Cubs,Ibata_2007_M31_M33_Halos, McConnachie2008_Trio_LG_DGs,Richardson_2011_PAndAS_Progeny}. Although detection limits constrain a complete sample of M31 dwarf galaxies \citep{Doliva-Dolinsky_2022}, these discoveries have facilitated comparative analyses between the MW and M31 satellite systems, clearly highlighting their differences and enriching our understanding of galaxy evolution in varied cosmic environments. M31 satellites are distributed asymmetrically, with the majority of them lying on the Andromeda galaxy's near side \citep{Savino2022}, and it has been proposed that half of the satellites lie on a thin plane \citep{Ibata2013}. M31 satellites display more extended star formation histories (SFHs) than their MW counterparts \citep[e.g.,][]{Weisz_2014_Ext_SHFs_AndII_AndXVI, Skillman_2017_ISLAndS_SFHs_6_M31_dSphS,Weisz2019, McQuinn2023_PegasusW,Savino2023_SFH}, suggesting that satellite evolution may be influenced by the host's environment \citep{Weisz2019, Savino2023_SFH}. These more extended SFHs are reflected also in the redder horizontal branches (HBs) observed in M31 satellites \citep[e.g., ][]{DaCosta_1996_AndI,DaCosta_2000_AndII, DaCosta_2002_AndIII, McConnachie_2007_SuprimeCam_AndII, Martin_2017_And_redHB, Jennings_2023_UFD_HB_Ages}, compared to those in MW UFDs, indicating episodes of star formation after reionization, suggesting that the impact of reionization is not uniform across UFDs \citep{Skillman_2017_ISLAndS_SFHs_6_M31_dSphS, Weisz2019, Savino2023_SFH}. To enable more direct comparisons between these satellite systems, continued efforts to discover and characterize the faint UFD population of M31 are essential. 

Building on this need, this paper reports the discovery of Andromeda XXXV (And XXXV), a new ultrafaint stellar system in M31's halo, identified as an overdensity of stars within the PAndAS footprint (Figure \ref{fig:PAndAS_Dens_Map}). We adopted the name And XXXV because of its association with M31, given that the system's size and luminosity are consistent with the known population of UFDs in the Local Group, and given indications in its stellar populations of complex stellar populations like those of other M31 dwarf galaxy satellites. We detail the detection method in Section \ref{sec:detection}, and Section \ref{sec:photometry} discusses the Hubble Space Telescope (HST) follow-up observations and the data reduction, and presents the color--magnitude diagram (CMD) of And XXXV. Section \ref{sec:strc_params} describes the derivation of the structural parameters. The methods used to determine the distance are outlined in Section \ref{sec:distances}. In Section \ref{sec:CMD_metallicity} we estimate the mean metallicity of And XXXV. We continue by discussing how we measure the luminosity and stellar mass of the new system in Section \ref{sec:luminosity}. Finally, the last section of the paper summarizes and discusses our results, and places And XXXV into context with known systems in the Local Group. 

\section{Discovery of Andromeda XXXV}\label{sec:detection} 
And XXXV was discovered by our search algorithm (detailed below) as an overdensity of stars around M31 in the PAndAS survey. The PAndAS survey is a Large Programme of the Canada--France--Hawaii Telescope (CFHT) conducted from 2008 to 2011 with imaging undertaken by the MegaPrime/MegaCam wide-field camera (see \citealt{Ibata_2007_M31_M33_Halos, McConnachie2008_Trio_LG_DGs, McConnachie2018, Richardson_2011_PAndAS_Progeny} for more details). With a coverage of $\sim$390\,deg$^2$, the survey mapped a region within a projected radius of $R_{\text{proj}}\lesssim150$\,kpc from M31's center and $R_{\text{proj}}\sim50$\,kpc from Triangulum (M33)'s center (Figure \ref{fig:PAndAS_Dens_Map}). To search for satellites within the PAndAS footprint, we use the `Combined stellar catalog' from \citet{McConnachie2018}, adopting sources that are stellar (with morphological classifications in {\it both} bands of ``$-1$" or ``$-2$"; \citealt{McConnachie2018}), with $i<24$, and are selected to have colors and magnitudes similar to metal-poor red giant branch (RGB) stars (a CMD selection sloped along the metal-poor RGB with width 0.3\,mag, sensitive to stars more metal-poor than $[\text{Fe}/\text{H}]\sim -1.3$). Assuming a distance to M31 of $D_{\odot}=776^{+22}_{-21}$\,kpc \citep{Savino2022}, we select three trial distances for study: 0.5\,mag closer in distance modulus from M31, M31's distance, and 0.5\,mag more distant from M31.

We identify overdensities of RGB stars on 100--800\,pc scales following \citet{Bell2022}. The goal is to identify statistically significant clumps of metal-poor RGB star candidates on small scales over and above the larger-scale density field from M31's massive stellar halo and prominent streams. We perform kernel density estimation using a top-hat kernel with radii of 100--800\,pc, sampling the distribution every 50\,pc. In order to be called an overdensity, we require a Poisson probability of being drawn from the spatially varying background (assessed using a 4\,kpc top-hat radius) of $P<10^{-6}$, excluding an ellipse with major axis 35\,kpc and minor axis 17.5\,kpc around M31. We then determine a more tailored measure of significance by allowing up to a 400\,pc shift in the center and choosing the best significance in a range of apertures between 100 and 800\,pc. Candidates are those objects that have five or more stars and a final probability $P<10^{-7}$ of being drawn from the background by chance alone.  

At this stage, hundreds of overdensities of candidate metal-poor RGB stars are found. All known satellites and a significant number of the known star clusters in the PAndAS footprint were recovered \citep{Huxor_2014_final_pandas_catalogue, McConnachie2018}. We discard candidates that were identified by \citet{Martin2013} and classified as contaminants---this includes unresolved globular clusters around background relatively nearby elliptical galaxies ($z\sim$ 0.03--0.1). Other candidates are visually inspected. Some candidates are clearly associated with background galaxy clusters---the ``blue RGB star" candidates are unresolved, very faint galaxies in these clusters.   

Of those candidates that remain, And XXXV is a clear, relatively compact stellar system---more extended than a typical globular cluster, with several RGB stars being detected by PAndAS, but smaller than $r\sim 100$\,pc in size. As far as we can tell, this stellar concentration is not in existing catalogs\footnote{We do not speculate as to why And XXXV was not recovered in earlier catalogs; it is fainter and smaller than known M31 dwarf galaxies \citep{Martin2009_Pandas_Cubs, Martin2016, McConnachie2018, Savino2022}, and is faint and very extended compared to known M31 globular clusters \citep{Huxor_2014_final_pandas_catalogue}.}. The location of And XXXV within the PAndAS footprint can be seen in the left panel of Figure \ref{fig:PAndAS_Dens_Map}. The upper right panel of Figure \ref{fig:PAndAS_Dens_Map} also includes the PAndAS CFHT/MegaCam $g$-band discovery mosaic of And XXXV.

\begin{deluxetable*}{@{\extracolsep{\fill}}lccc}
\tablewidth{0pt} 
\tablecaption{Properties of Andromeda XXXV. \label{table:properties}}
\tablehead{\colhead{\hspace{-1.7cm} Parameter} & \colhead{Description} &  \multicolumn2c{Value} }
\startdata
HST Name & ... &  \multicolumn2c{OBJ-002638+400622}  \\
\hline 
& &  {\bf Exponential} & {\bf Plummer}\\
$\alpha$ (J2000) & R.A. & $00^{\text{h}}26^{\text{m}}38\fs6 \; ^{+1.7}_{-1.6} \, \arcsec$ & $00^{\text{h}}26^{\text{m}}38\fs6 \; ^{+1.7}_{-1.8} \, \arcsec$ \\ 
$\delta$ (J2000) & Decl. & $+40^{\circ}6 \arcmin 29\farcs3  \; ^{+2.8}_{-2.8} \, \arcsec$  & $+40^{\circ}6 \arcmin 29\farcs2 \; ^{+3.0}_{-3.0} \, \arcsec$ \\
$\theta$ ($^{\circ}$E of N) & Position angle & $80 \,^{+16}_{-16}$  & $82 \, ^{+17}_{-17}$ \\
$\epsilon$ & Ellipticity & $ 0.43 \, ^{+0.16}_{-0.22}$  & $0.43 \, ^{+0.17}_{-0.23} $ \\
 & ($\epsilon = 1-b/a$) &  &\\
$a_h$ (arcmin) & Angular elliptical& $0.26\,^{+0.08}_{-0.06}$ & $0.27 \,^{+0.09}_{-0.06}$\\
& Half-light radius &  & \\
$a_h$ (pc) $^{\text{a}}$& Physical elliptical  & $71\,^{+21}_{-16}$ & $72\,^{+24}_{-17}$\\
 & Half-light radius & &\\
$r_h$ (arcmin) & Angular circularized & $ \, 0.20\,^{+0.05}_{-0.04}$ & $ \, 0.20\,^{+0.05}_{-0.04}$ \\
 & Half-light radius &  & \\
$r_h$ (pc) $^{\text{a}}$ & Physical circularized & $ 53\,^{+13}_{-11}$ & $54\,^{+15}_{-12}$\\
 & Half-light radius &  & \\
$N_{\ast}$  & Number of stars & $ 22\,^{+3}_{-3}$ & $23\,^{+3}_{-4}$\\
$\Sigma_{\,\text{b}}$ (stars arcmin$^{-2}$)  & Background density & $1.5$ & $1.4$\\
$\Delta \text{BIC}^{\text{ b}}_{\text{circ}|\text{ellip}}$ & BIC $^{\text{c}}$ score difference & $1.2$ & $ 6.3$ \\ 
$\Delta \text{BIC}^{\text{ d}}_{\text{Plummer}|\text{Exp}}$ & BIC $^{\text{c}}$ score difference & \multicolumn2c{ $1.3$} \\
\hline 
$\mu_{0}$ (mag) & Distance modulus & \multicolumn2c{$24.83\,^{+0.17}_{-0.15}$} \\
$D_{\odot}$ (kpc) & Heliocentric distance & \multicolumn2c{$927\,^{+76}_{-63}$ }   \\
$D_{\,\text{M31}}$ (kpc) $^{\text{a}}$ & Distance from M31  &  \multicolumn2c{ $158\,^{+57}_{-45}$} \\ 
$M_{V}$ (mag) $^{\text{a}}$ &  Absolute $V$-band magnitude& \multicolumn2c{ $-5.2\, \pm 0.3$}\\
$L$ $(L_{\odot})$ $^{\text{a}}$ & Luminosity & \multicolumn2c{ $(1.0\,^{+0.3}_{-0.2})\times 10^{4}$}\\ 
$\mu_V$ (mag arcsec$^{-2}$) & Surface brightness & \multicolumn2c{$27.0\,\pm 0.4$}\\
$M_{\ast}$ $(M_{\odot})$ $^{\text{a}}$& Stellar mass & \multicolumn2c{$(2.0\, \pm 0.4)\times 10^{4}$}\\
$\langle [\text{M}/\text{H}]\rangle$ (dex) $^{\text{a, e}}$  & Mean metallicity & \multicolumn2c{$\sim -1.7\, \pm 0.3$} \\
$\langle [\text{Fe}/\text{H}]\rangle$ (dex) $^{\text{a, e, f}}$ & Mean iron abundance & \multicolumn2c{$\sim -1.9\, \pm 0.3$} \\
\hline
\enddata
\tablecomments{(a) Distance dependent property; distance uncertainties were propagated through all properties except through the physical elliptical and azimuthally averaged half-light radii. (b) $\Delta \text{BIC}_{\text{circ}|\text{ellip}} = \text{BIC}_{\text{circular}} - \text{BIC}_{\text{elliptical}}$ where a positive number means an elliptical profile is favored over a circular profile. The BIC scores for the circular versions of the models were evaluated setting $\epsilon =0$ and $\theta = 0 \, ^{\circ}$. (c) BIC denotes Bayesian information criterion. (d) $\Delta \text{BIC}_{\text{Plummer}|\text{Exp}} = \text{BIC}_{\text{Plummer}} - \text{BIC}_{\text{Exp}}$ where a positive number means the exponential profile is favored over the Plummer profile. (e) Photometric. (f) The mean iron abundance was estimated assuming $[\alpha/\text{Fe}]=+0.3$ and using the relation from \citet{Salaris_1993_alpha_iron}. } 
\end{deluxetable*}

\section{HST Follow-up Observations}\label{sec:photometry}
We conducted follow-up imaging from HST's Wide Field Camera 3 (WFC3) in the F606W and F814W UVIS filters as part of HST SNAP 17158 (PI: Bell).\footnote{The HST data presented in this paper are available through the Mikulski Archive for Space Telescopes at the Space Telescope Science Institute, \dataset[doi:10.17909/mkz8-z964]{https://doi.org/10.17909/mkz8-z964}.} The field of view (FOV) is $162 \arcsec \times 162 \arcsec$, with a plate scale of $0\farcs04$\,pix$^{-1}$. We adopted a visit design that obtained $2\times150$\,s F606W exposures (to allow for cosmic-ray rejection) and one 350\,s exposure in F814W. This visit design has the disadvantage that some area is lost to cosmic rays in F814W, but it keeps the SNAP visit as short as possible to maximize the chance of being scheduled for execution. The lower right panel of Figure \ref{fig:PAndAS_Dens_Map} presents the HST/WFC3 F606W follow-up image of And XXXV.

We reduced the data and performed PSF-fitting photometry using the software package \texttt{DOLPHOT} \citep{dolphot2000, dolphot2016} using TinyTim PSFs \citep{krist2011_tinytim}. In order to isolate sources that are likely to be stars uncontaminated by cosmic rays, we select for further consideration unmasked sources that satisfy the following criteria: the signal-to-noise ratio is $\geqslant$5 in each filter, the square of the sharpness parameter in the two passbands is $<$0.1, the sum of the crowding parameters in the two bands is $<$1.0, the error flag is $\leqslant$3 in each passband, and the object type is $\leqslant$2. We apply manual aperture corrections following the method described in \citet{Jang_2021} and \citet{Jang_2023}.

We also ran a suite of artificial star tests (ASTs) in order to quantify the completeness and photometric uncertainties of our sample. Around 270,000 artificial stars were drawn from a uniform grid of sources with magnitudes between $22<\text{F814W}<28$ and colors between $-0.5<\text{F606W}-\text{F814W}<2.5$ and injected across the field of view. We applied the same quality metrics to recovered sources to derive the observational effects of the HST/WFC3 image for each filter. The 50\% (80\%) completeness limit of our HST data is $\text{F814W}= 25.31 \, (24.50)$\,mag and $\text{F606W}= 26.29 \, (25.56)$\,mag. 

The final photometric catalog was corrected for Galactic foreground extinction using the maps and calibrations provided by \citet{Schlegel_1998} and \citet{Schlafly_2011}. The left panel of Figure \ref{fig:Cand1_Diag_Dens_plot} displays the extinction-corrected CMD for stars within the HST/WFC3 field of view that meet our photometric cuts. Stars within two half-light radii (as derived in Section \ref{sec:strc_params}) are color coded according to their elliptical radius from And XXXV's center. Old PARSEC isochrones \citep{2012_Parsec} at varying metallicities are overlaid and shifted to a distance of $927$\,kpc (estimated in Section \ref{sec:distances}), highlighting a sparsely populated RGB and a red HB (RHB) feature around $\text{F814W}_0 \sim 24.6$\,mag. We discuss the main properties of And XXXV in the next section. 

\section{Properties of Andromeda XXXV}
\subsection{Structure} \label{sec:strc_params}
We determined the structural properties of And XXXV using a Markov Chain Monte Carlo (MCMC) maximum likelihood algorithm, following the methodology outlined by \citet{Martin2008, Martin2016}, and \citet{McQuinn2023_PegasusW}. We employ the \texttt{emcee} package \citep{Foreman-Mackey2013} to sample the probability density functions (PDFs) of the structural parameters, focusing on CMD-selected stars defined within a selection box (left panel of Figure \ref{fig:Cand1_Diag_Dens_plot}) tailored around old, relatively metal-poor PARSEC isochrones \citep{2012_Parsec}. To minimize incompleteness effects, we include only stars exceeding our 80\% completeness limits ($\text{F814W}_0 \leqslant 24.50$\,mag and $\text{F606W}_0 \leqslant 25.56$\,mag) and with colors $\text{F606W}_0-\text{F814W}_0 < 1.4$\,mag.

We model the spatial distribution of stars with two radial density profiles: an exponential profile and a Plummer profile \citep{Plummer1911}. Each model is defined as a function of elliptical radius with a central stellar density and a constant background term. Both models each return six parameters: the centroid of the stellar distribution $(x_0, y_0)$, the ellipticity $\epsilon$ (defined as $\epsilon = 1 - b/a$ with $b/a$ representing the minor-to-major axis ratio), the position angle of the major axis $\theta$ in degrees east of north, the total number of member stars $N_{\ast}$ within the color-magnitude space, and the elliptical half-light radius $a_h$ (which represents the projected semimajor axis that encompasses half of the total integrated surface density;\footnote{The exponential scale length $r_e$ is related to the elliptical half-light radius via $a_h = 1.68 r_e$, and the Plummer scale length $r_p$ is equivalent to $a_h$ ($r_p=a_h$). The circularized half-light radius $r_h$ is related to the elliptical half-light radius $a_h$ via the equation $r_h = a_h \sqrt{1-\epsilon}$.} \citealt{Drlica-Wagner2015, Richstein2022}). We implement flat uniform priors for all the structural parameters and execute a total of 15,000 iterations for each model, with $\sim$2000 initial steps discarded as burn-in.  

The MCMC analysis achieved satisfactory convergence for all parameters. We present the medians of the parameters as the best-fit values, with their uncertainties defined by the 16th and 84th percentiles. Comprehensive results, including final parameter estimates and derived properties, are detailed in Table \ref{table:properties} and visualized in Figure \ref{fig:Cand1_corner_plots}, which displays the corner plots of the posterior distributions and correlations of the structural parameters. The middle panel of Figure \ref{fig:Cand1_Diag_Dens_plot} shows the spatial distribution of all point sources meeting our photometric criteria, overlaid with ellipses at $2 \times a_h$ and $3\times a_h$ that represent the best-fit exponential and Plummer models, while the bottom right panel includes the binned radial density profiles of And XXXV, with the data binned in elliptical annuli according to the best-fit centroid, ellipticity, and position angle of each model. 

We evaluate the Bayesian information criterion (BIC) to determine how well each model captures the stellar distribution of And XXXV. A lower BIC score indicates a more suitable model. The BIC scores for the Plummer and exponential profiles are $\text{BIC}_{\text{Plummer}} = 70.1$ and $\text{BIC}_{\text{Exp}} = 68.8$, respectively. Additionally, when assessing circular versions of these models (with $\epsilon = 0$ and $\theta = 0$), the BIC scores increased (see Table \ref{table:properties}), suggesting that the elliptical profiles are better representations of the data. Given that the exponential model is a slightly better fit and the widespread use of the exponential profile in nearly all literature studies, we will base our subsequent analysis on the structural parameters derived from this model.

\subsection{Stellar Populations and Distance}\label{sec:Stellar_pop_Distances}
\begin{figure}[hbt!]
\centering
\includegraphics[width=1.0\columnwidth,]{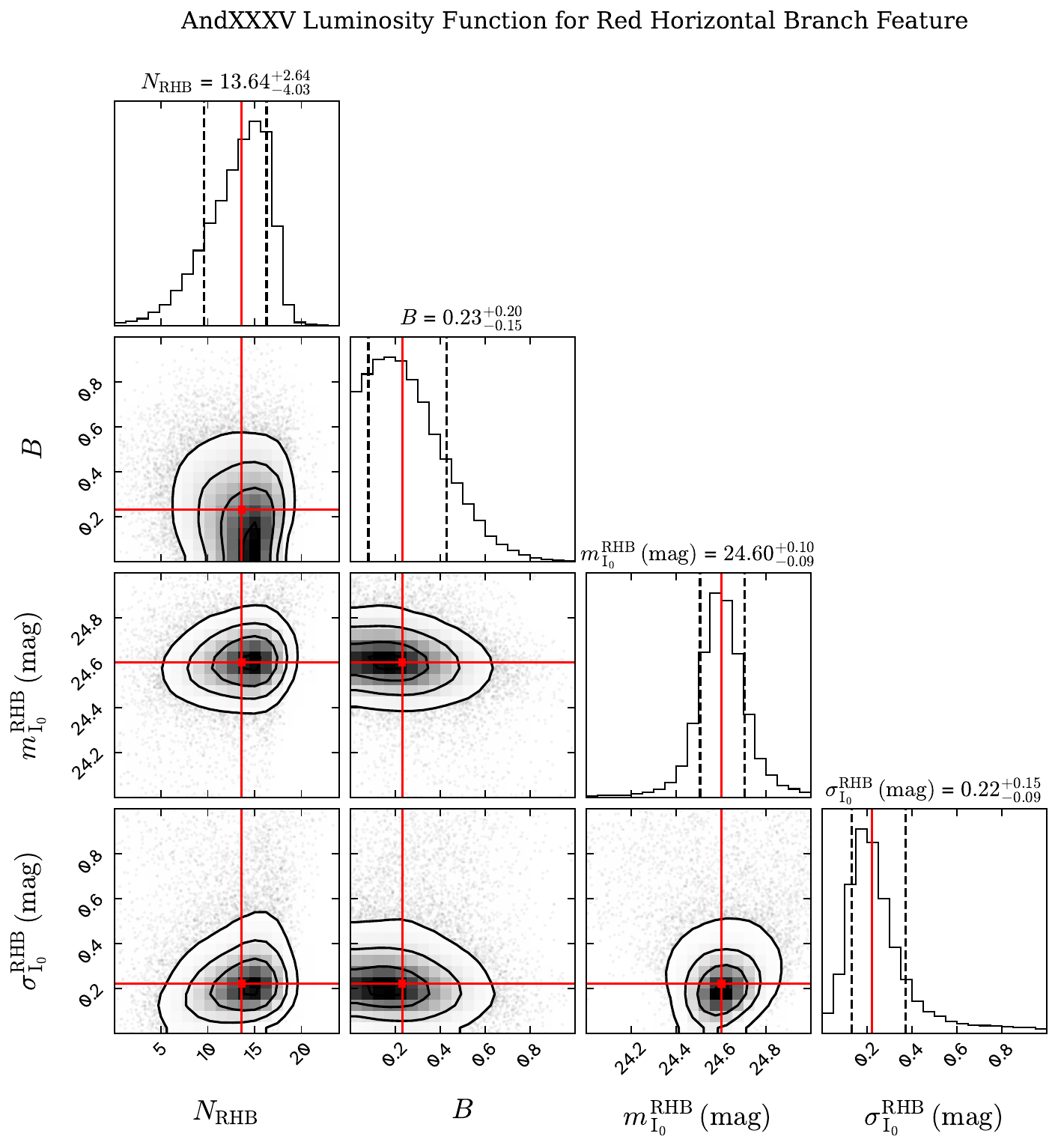}
\caption{Corner plots displaying the posterior distributions of the $I$-band luminosity function model for the RHB feature, including the apparent magnitude of the RHB ($m_{I_0}^{\text{RHB}}$), its width ($\sigma_{I_0}^{\text{RHB}}$), the number of candidate RHB stars ($N_{\text{RHB}}$), and the exponential slope ($B$). The posterior distributions are based on over 10,000 MCMC iterations after burn-in. The red solid lines and the dashed gray lines represent the medians and the 16th and 84th percentiles of each parameter, respectively.\label{fig:Cand1_corner_LF_plots}}
\end{figure}
We base our distance estimate for And XXXV on the RHB feature in its CMD, since it has too few stars to define a clear tip of the RGB (TRGB; see left panel of Figure \ref{fig:Cand1_Diag_Dens_plot}). RHB stars are core-helium-burning stars that are slightly more metal-rich and likely younger than blue horizontal branch stars; these appear in a similar part of the CMD to typically more massive red clump (RC) stars. These stars are often observationally convenient standard candles \citep[e.g.,][]{Alves_2002_Hipparcos_RC}, but their luminosity depends weakly on stellar populations, requiring joint consideration of both the full CMD information and RHB feature specifically \citep{Girardi_Salaris_2001, Monachesi_2011_M32, Girardi_2016_RC_stars}.

In particular, the CMD for And XXXV shows both blue metal-poor RGB stars (signaling old and metal-poor stars) and RHB stars (suggesting slightly younger and metal-richer stars; \citealt{Martin_2017_And_redHB,Weisz2019}). This is important---And XXXV's CMD directly suggests a range of stellar metallicities (and ages), supporting its classification as Andromeda's faintest dwarf galaxy companion. While our CMD is too sparse to attempt detailed SFH fitting, analyses of deeper, better-populated, similar-looking CMDs for other Andromeda satellites \citep[e.g.,][]{Weisz2019,Savino2023_SFH} yields SFHs that continued to form stars as recently as $\sim$6\,Gyr ago. We use this picture to inform our RHB-based distance estimates for And XXXV. 

\subsubsection{Distance}\label{sec:distances} 
To determine the distance to And XXXV, we first transform our $\text{F606W}_0$ and $\text{F814W}_0$ magnitudes into Johnson $V_0$ and $I_0$ magnitudes. This transformation is performed iteratively, following \citeauthor{Sirianni_2005} (\citeyear{Sirianni_2005}; their Equation (12) and coefficients from Table 22), and continued until numerical convergence is achieved. We then select sources within $0.4 < (V - I)_0 < 1.25$ and $21.8 < I_0 < 25.6$ in the transformed $I_0 - (V - I)_0$ CMD.

We proceed to measure the apparent magnitude, $m_{I_0}^{\text{RHB}}$, and the width, $\sigma_{I_0}^{\text{RHB}}$, of the RHB/RC of And XXXV using a MCMC maximum likelihood approach that fits a luminosity function (LF) to the observed distribution of likely RHB stars \citep[e.g.,][]{Nataf_2013_RC_reddening, Cohen_2018_RC_GSS}. The LF consists of a Gaussian component to model the RHB and an exponential to represent the RGB background (first two terms of Equation (1) from \citealt{Cohen_2018_RC_GSS}). We define the standard deviation of the RHB Gaussian component as the quadrature sum of its intrinsic width, $\sigma_{I_0}^{\text{RHB}}$, and photometric uncertainties, $\sigma_{I_0}^{\text{phot}}$, derived from exponential fits to our AST results \citep[e.g.,][]{ Nataf_2013_RC_reddening, Monachesi2016, Cohen_2018_RC_GSS}. We employ flat priors for all model parameters. Additionally, we constrain the number of candidate RHB stars, $N_{\text{RHB}}$, and RGB background stars to not exceed the total number of stars within the color-magnitude region. Figure \ref{fig:Cand1_corner_LF_plots} displays the posterior distributions for $m_{I_0}^{\text{RHB}}$, $\sigma_{I_0}^{\text{RHB}}$, $N_{\text{RHB}}$, and the slope of the exponential, $B$. In the top right panel of Figure \ref{fig:Cand1_Diag_Dens_plot}, we show the $I$-band LF around the RHB feature along with the best-fit model.

Because And XXXV's CMD is sparsely populated, the inclusion or exclusion of CMD-selected sources can influence the best-fit parameters of the LF model \citep[e.g.,][]{McQuinn2024_LeoK_LeoM}. To account for this, we iterate over the color-magnitude region and repeat the maximum likelihood approach multiple times, estimating an additional systematic uncertainty associated with data selection bias. Incorporating both MCMC and resampling uncertainties, the best-fit apparent magnitude and the width of the RHB are $m_{I_0}^{\text{RHB}}= 24.60\,^{+0.10\,+0.06}_{-0.09 \,-0.05} $\,mag and $\sigma_{I_0}^{\text{RHB}}= 0.22 \,^{+0.15\,+0.16}_{-0.09\,-0.03}$\,mag, respectively, with the LF model predicting $N_{\text{RHB}} = 14\,^{+3 \,+1}_{-4\,-1}$ candidate RHB stars.
 
We calculate the distance modulus using Equation 8 from \citet{Girardi_2016_RC_stars}, $\mu_0 = m_{I_0}^{\text{RHB}}-M_{I , \; \text{local} }^{\text{RHB}}+\Delta M_{I}^{\text{RHB}}$, where the ``population correction" term, $\Delta M_{I}^{\text{RHB}} = M_{I , \; \text{local} }^{\text{RHB}} - M_{I, \; \text{And XXXV} }^{\text{RHB}} $, quantifies the difference between the RHB absolute magnitude in the local star sample and in And XXXV due to different SFHs \citep{Girardi_Salaris_2001, Girardi_2016_RC_stars}. Several studies indicate that $M_{I}^{\text{RHB}}$ is influenced by age and metallicity \citep[e.g.,][]{Girardi_Salaris_2001, Salaris_2002_RC_PopEffects, 2010_Pietrzynski__Pop_Effects_IVband, Girardi_2016_RC_stars}, making it useful to approximately account for these effects using population synthesis models \citep{Girardi_Salaris_2001}. 

Inspired by SFH fits to more luminous Andromeda satellites with deep CMD data (e.g., \citealt{Savino2023_SFH}), we create a model population with a strongly declining star formation rate (SFR) between 12.5 and 8\,Gyr ago, with metallicity increasing from $[\text{M}/\text{H}] = -2.2$\,dex to  $[\text{M}/\text{H}] = -0.8$\,dex in that time. The resulting sparse model CMD is similar to the observed one. In particular, an RHB is present, stemming primarily from the subset of stars that are somewhat younger at $\sim$10\,Gyr old, and metal-richer at $[\text{M}/\text{H}] \sim -1$\,dex, than average. This age-metallicity combination yields $M_{I,\;\text{And XXXV}}^{\text{RHB}} = -0.237\,^{+0.028}_{-0.015} $\,mag based on the theoretical core-helium-burning models of \citet{Girardi_Salaris_2001}. The approximate uncertainties in $M_{I,\;\text{And XXXV}}^{\text{RHB}}$ are estimated by considering slightly different ages and metallicities that produce plausible-looking CMDs. When compared with the $M_{I,\;\text{local} }^{\text{RHB}}$ for local Hipparcos RHB/RC stars, $M_{I,\;\text{local}}^{\text{RHB}} = -0.26 \pm 0.03$\,mag \citep{Alves_2002_Hipparcos_RC}, this results in a population correction of $\Delta M_{I}^{\text{RHB}} = -0.023$\,mag.

The resulting distance modulus for And XXXV is $\mu_0 = 24.83 \,^{+0.17}_{-0.15}$\,mag, corresponding to a heliocentric distance of $927\,^{+76}_{-63}$\,kpc. The uncertainties include contributions from the MCMC LF fit and data selection bias ($^{+0.16}_{-0.14}$\,mag), the theoretically derived $M_{I,\;\text{And XXXV}}^{\text{RHB}}$ ($^{+0.028}_{-0.015}$\,mag), and the RHB calibration ($\pm \, 0.03$\,mag). Based on this distance and the best-fit exponential model, the physical elliptical and azimuthally averaged half-light radii of And XXXV are $a_h = 71\,^{+21}_{-16}$\,pc and $r_h = 53\,^{+13}_{-11}$\,pc, respectively. Positioned at a three-dimensional distance of $D_{\,\text{M31}} = 158\,^{+57}_{-45}$\,kpc from the center of M31, and well within M31's virial radius of $266$\,kpc \citep{Fardal_2013_M31_Mass, Putman_2021_Gas_LG}, we confirm the status of And XXXV as a companion of Andromeda. 

\subsubsection{Metallicity and Stellar Populations}\label{sec:CMD_metallicity}
As previously noted, the CMD of And XXXV shows a sparse, vertical sequence of stars at magnitudes brighter than $\text{F814W}_0 \sim 24.5$\,mag and within a color range of $0.5 < \text{F606W}_0-\text{F814W}_0 < 1.2$\,mag. At the distance modulus of $\mu_{0} = 24.83\,^{+0.17}_{-0.15}$\,mag, these stars are consistent with belonging to the RGB. We see very little evidence for stars bluer than an ancient RGB, suggesting a lack of star formation in the last Gyr. The RGB has some width in excess of estimated uncertainties; when considered in concert with the RHB of And XXXV, this further supports the idea of a spread in ages and metallicities in this dwarf galaxy. 

To estimate the mean photometric metallicity of And XXXV, we employ two methods based on the color-magnitude properties of these RGB stars. Color-based metallicity estimates rely on model calibration or are calibrated to specific systems (e.g., Galactic globular clusters), and depend on the foreground Galactic extinction. The first method compares the color of RGB stars to isochrones, following \citet{Ogami_2024_met}. 
We use radial basis function (RBF) interpolation from Python's \texttt{SciPy} package to match the CMD positions of each RGB star to a grid of 12.5\,Gyr old PARSEC isochrones \citep{2012_Parsec} ranging from $-2.2 < [\text{M}/\text{H}] < 0.5$\,dex, deriving a mean photometric metallicity of $\langle [\text{M}/\text{H}] \rangle \sim -1.5\,\pm 0.3$\,dex---similar to the mean metallicity of the composite stellar population used to determine the RHB population correction for And XXXV described above. The second method uses the relationship between the RGB slope and the metallicity; this has the advantage of being independent of foreground Galactic extinction uncertainties. \citet{Streich_2014_RGB_met} relates the slope of the RGB between the level of the HB and two magnitudes brighter to the metallicity of Galactic globular clusters; on this basis, we estimate a metallicity $\langle [\text{M}/\text{H}] \rangle \sim -1.9\,\pm 0.6$\,dex, including errors from both the RGB slope-metallicity relation and the RGB slope itself.

Assuming a global $\alpha$-element enhancement of $[\alpha/\text{Fe}]=+0.3$\,dex and using the relation from \citet{Salaris_1993_alpha_iron}, the mean iron abundances derived from the positions of RGB stars in the CMD and the RGB slope method are $\langle [\text{Fe}/\text{H}] \rangle \sim -1.7\,\pm 0.3$\,dex and $\langle [\text{Fe}/\text{H}] \rangle \sim -2.1\, \pm 0.6$\,dex, respectively. Averaging the two $[\text{M}/\text{H}]$ estimates, we obtain $\langle [\text{M}/\text{H}] \rangle \sim -1.7\,\pm 0.3$\,dex and $\langle [\text{Fe}/\text{H}] \rangle \sim -1.9\,\pm 0.3$\,dex. 

For visual reference, we overplot 12.5\,Gyr PARSEC \citep{2012_Parsec} isochrones with $[\text{M}/\text{H}] = -1.9$ and $-1.5$\,dex, and a 10\,Gyr isochrone with $[\text{M}/\text{H}] = -1.0$\,dex at the derived distance to And XXXV onto its CMD in the left panel of Figure \ref{fig:Cand1_Diag_Dens_plot}. We note that there are three member stars, found at $\text{F814W}_0\sim$ 20.0--21.0\,mag, as bright as or brighter than the expected And XXXV TRGB. One of these stars (at $\text{F814W}_0\sim$20.7\,mag) is near the elliptical half-light radius of And XXXV and aligns with the TRGB of the most metal-poor isochrone ($[\text{M}/\text{H}]= -1.9$\,dex). Thus, it is possible that this star corresponds to the TRGB at our derived distance of $D_{\odot}=927\,^{+76}_{-63}$\,kpc. The other two bright stars (around $\text{F814W}_0 \sim 20.3$\,mag) are farther out at elliptical radii of $R_{e} \sim$ 1.6--1.8\,$a_h$, above the expected TRGB. While they could, in principle, be And XXXV asymptotic giant branch stars, this would be extremely unusual given how sparsely populated the TRGB is. Instead, we suggest that they may be M31 halo members, as they align well with the expected TRGBs ($\text{F814W}_0 \sim 20.3$\,mag) of metal-poor isochrones at M31's distance \citep[$\mu_{0} = 24.45\,\pm 0.06$\,mag; ][]{Savino2022}; it is also possible that they might be MW foreground stars. 

\begin{figure*}[htb!]
\centering
\includegraphics[width=0.48\textwidth]{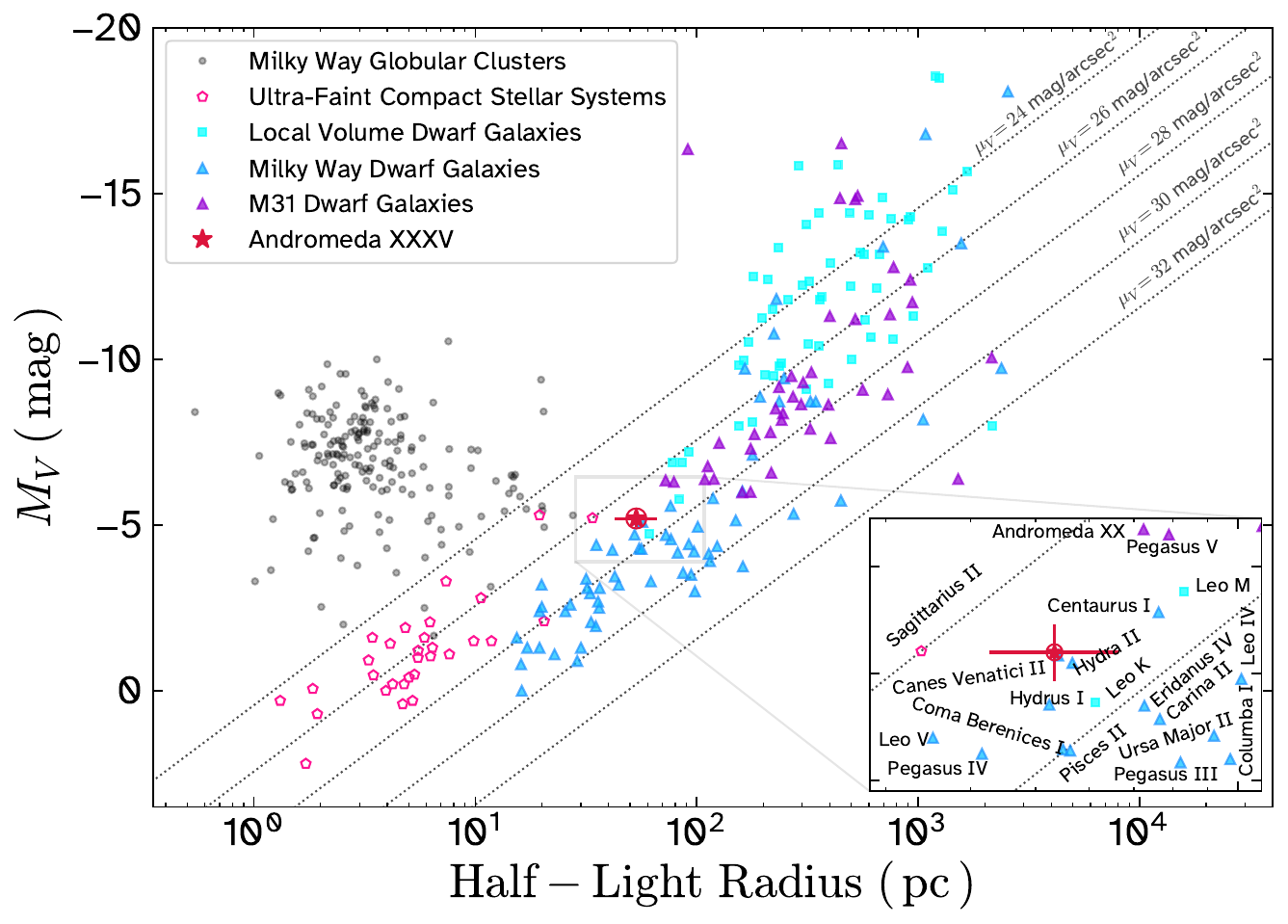}
\includegraphics[width=0.49\textwidth]{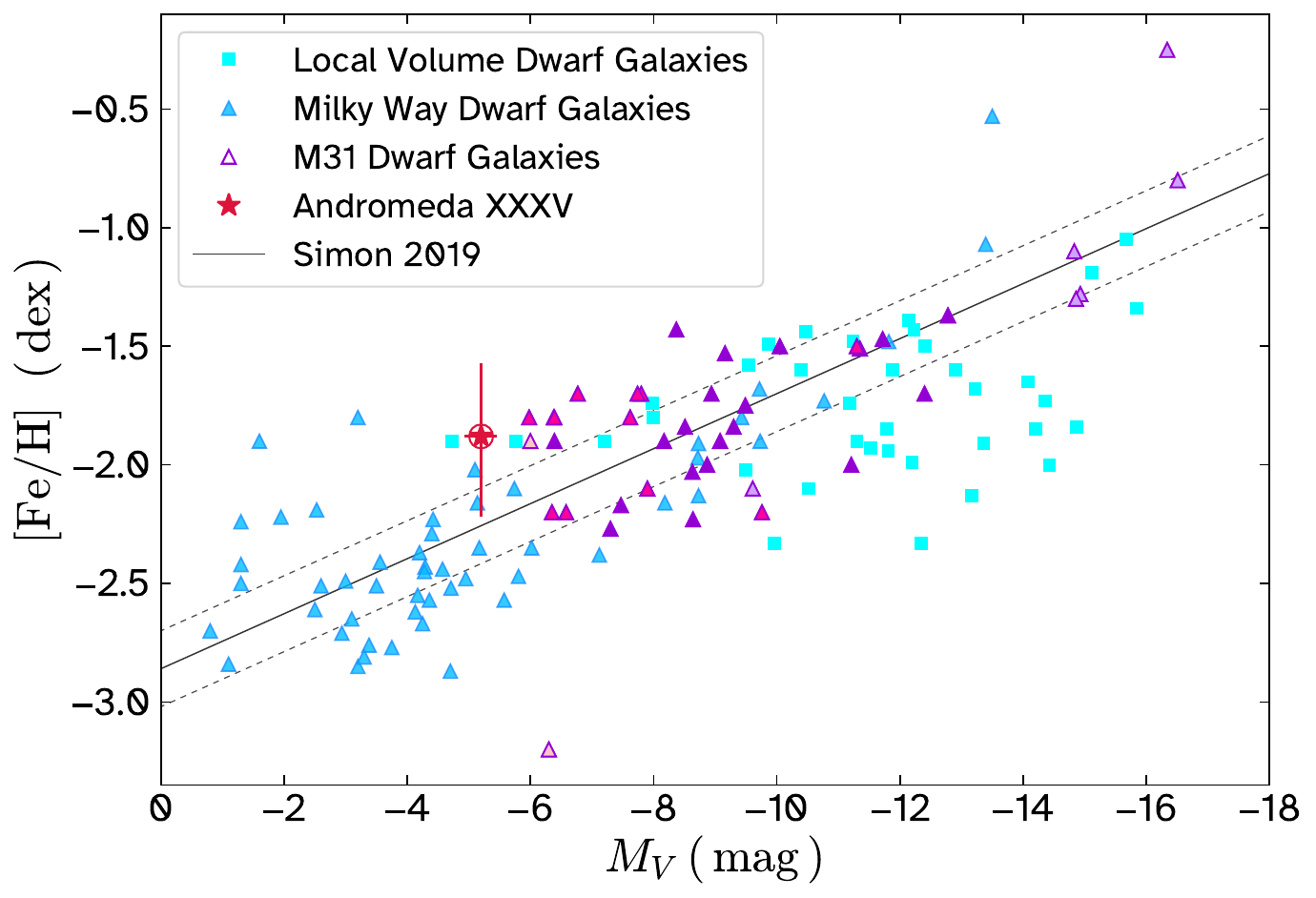}
\caption{Left: absolute $V$-band magnitude ($M_V$) vs. azimuthally averaged physical half-light radius ($r_h$) for various stellar systems in the Local Group, including MW classical globular clusters, MW ultrafaint compact stellar systems (ambiguous objects), both MW and M31 dwarf galaxies, and other Local Volume dwarf galaxies. An inset highlights systems comparable to And XXXV in the luminosity--size plane. Data for all systems except And XXXV are sourced from the Local Volume Database \citep{Pace_2024_LVDB}, and references therein. Right: metallicity--luminosity relation for Local Group dwarf galaxies, with the black solid line representing the best-fit relation from \citet{Simon2019}. The dashed lines represent a $1\sigma$ scatter of $0.16$\,dex around the fit.  M31 dwarf galaxies are outlined in violet triangles (filled dark pink triangles, taken from \citealt{Collins_2013_Kin_Study_M31_Sats}; filled light violet triangles, taken from \citealt{McConnachie2012}, and references therein; filled light pink triangles, taken from \citealt{Collins_2022_PegV} and \citealt{Collins_2024_Pisces_VII}). And XXXV is marked by a filled red star with error bars. 
Other measurements, including those for MW and Local Volume dwarfs, are taken from the Local Volume Database \citep{Pace_2024_LVDB}, and references therein. 
\label{fig:Luminosity_size_metallicity}}
\end{figure*}

\subsection{Luminosity and Stellar Mass}\label{sec:luminosity}
We derive the total luminosity of And XXXV using a method adapted from \citet{Martin2008, Martin2016}, following the implementation from the maximum-likelihood-based Ultrafaint GAlaxy LIkelihood toolkit (\texttt{ugali}, \citealt{Bechtol2015, Drlica-Wagner2020}). Assuming a Kroupa initial mass function (IMF; \citealt{Kroupa2002}), we generate a CMD PDF of the member stars of And XXXV using a 12.5 Gyr PARSEC isochrone \citep{2012_Parsec} with $[\text{M}/\text{H}] = -1.7$\,dex, and convolved with photometric uncertainties derived from ASTs. For each trial, the PDF isochrone is shifted by a random distance modulus, normally distributed around $24.83$\,mag with a standard deviation of $0.16$\,mag. Simultaneously, we draw a random number of stars, $N_{\ast i}$, from our MCMC-derived $N_{\ast}$ distribution. We then sample the CMD PDF until accumulating $N_{\ast i}$ stars falling within observational limits (randomly determined from our $\text{F606W}$ and $\text{F814W}$ completeness functions) and the CMD box used for deriving the structural parameters. Finally, we determine the total luminosity and stellar mass by summing the contributions from both observed and unobserved stars in each realization of the simulated stellar population. We repeat this process 1000 times to effectively propagate the uncertainties associated with the number of stars determined from our MCMC analysis and our distance measurement.

The total absolute $V$-band magnitude of And XXXV is $M_V=-5.2 \pm 0.3$, and its total stellar mass is  $M_{\ast}= (2.0\,\pm 0.4)\times 10^4\, M_{\odot}$, establishing it as the faintest known UFD of M31 to date; it is also the most compact known M31 UFD. Lastly, we determine the effective surface brightness by dividing half of the total flux by the area enclosed within one circularized half-light radius and converting to mag arcsec$^{-2}$, yielding $\mu_V= 27.0\, \pm 0.4$\,mag arcsec$^{-2}$. 

\section{Discussion and Conclusion} \label{sec:discussion}
We report the discovery of Andromeda XXXV, a new UFD in the halo of Andromeda, discovered in the PAndAS survey and confirmed with HST/WFC3 follow-up imaging. With its inferred heliocentric distance ($927^{+76}_{-63}$\,kpc) and three-dimensional separation from M31 ($D_{\,\text{M31}} = 158\,^{+57}_{-45}$\,kpc), And XXXV is established as a satellite of Andromeda. Like around half of M31's satellites, And XXXV lies off of M31's `plane of satellites' \citep{Ibata2013} at about $\sim$100\,kpc from the plane; we note also that And XXXV's distance places it on the far side of M31, in contrast to many of its other satellite galaxies (e.g., \citealt{Savino2022}).  

Being larger ($r_h = 53\,^{+13}_{-11}$\,pc) than all known MW and M31 star clusters---as far as we are aware, the current records are held by Sagittarius II (Sgr II) in the MW with $r_h \sim$34\,pc \citep{Longeard2021, Richstein2024} and PAndAS-33 in M31 with $r_h \sim$36\,pc \citep{Huxor_2014_final_pandas_catalogue}---and with a total $V$-band luminosity of $M_V=-5.2 \pm 0.3$, And XXXV aligns with other Local Group dwarf galaxies in the luminosity--size plane. The left panel of Figure \ref{fig:Luminosity_size_metallicity} shows the luminosity--size relationship for dwarf galaxies around the Local Group, and both MW globular clusters and ultrafaint compact systems\footnote{The properties of all systems except those of And XXXV were obtained from the Local Volume Database \citep{Pace_2024_LVDB}: \url{https://github.com/apace7/local_volume_database}.}. In the inset panel, systems with similar luminosity and half-light radius are highlighted. All but one of these systems are classified as galaxies. And XXXV shares a virtually identical structure and luminosity with Canes Venatici II (CVn II; $M_V \sim -5.2$, $r_h \sim 54$\,pc; \citealt{Munoz_2018_Str_of_MW_Sats}) and Hydra II (Hya II; $M_V \sim -5.1$, $r_h \sim 57$\,pc; \citealt{Munoz_2018_Str_of_MW_Sats, Richstein2024}), both of which are old, metal-poor ($[\text{Fe}/\text{H}] \sim -2.35$ and $-2.0$\,dex, respectively) MW companions \citep{Simon2019, Kirby_2015_HydraII_PiscesII}. The only system with ambiguous classification is Sgr II ($M_V \sim -5.2$, $r_h \sim 34$\,pc). Its velocity dispersion, $1.7\pm0.5$ km s$^{-1}$, is somewhat larger than expected for a globular cluster \citep[expected to be 0.5\,km s$^{-1}$; ][]{Baumgardt2022}, but its narrow metallicity distribution, $<0.2$\,dex, is consistent with Galactic globular clusters \citep{Longeard2021}. Its lack of mass segregation at $>$2$\sigma$ level (consistent with it being a galaxy) is also consistent with the expected degree of mass segregation if it were a star cluster \citep{Baumgardt2022}. Irrespective of Sgr II's classification, And XXXV is substantially larger in half-light radius than Sgr II, consistent with our classification of And XXXV as a galaxy.  

And XXXV's stellar populations are not particularly well constrained by our relatively shallow HST CMD. The CMD shows no evidence for young or intermediate age stars, but it shows evidence of an upper RGB and what appears to be a RHB. The RGB's color and slope are broadly consistent with a metallicity [M/H]$\sim -1.7\pm0.3$\,dex; assuming alpha enhancement, [Fe/H]$\sim -1.9\pm0.3$\,dex. This metallicity estimate is compared with the galaxy metallicity--luminosity relation in the right-hand side of Figure \ref{fig:Luminosity_size_metallicity}; it is within the scatter of the relation but somewhat on the high metallicity side for its luminosity. Considering that And XXXV's high metallicity relative to its luminosity could be a sign of tidal stripping, we estimate its tidal radius. We assume a line-of-sight velocity dispersion of $\sigma_{\text{los}} \sim 5$\,km s$^{-1}$ and an enclosed M31 halo mass of $M_{\text{enc}}(R \sim 160 \, \text{kpc}) \backsimeq 2.1 \times 10^{12} \, M_{\odot}$, derived from the circular rotational velocity $v_c = 240\pm38$\,km s$^{-1}$ reported by \citet{2010_Corbelli_M31_rot_curve}. With a half-light radius $r_h = 53\,^{+13}_{-11}$\,pc that is about 20$\times$ smaller than its estimated tidal radius $r_{t} \sim 1110$\,pc and a 3D distance from M31 of approximately 160 kpc, there is no argument strongly supporting that tidal stripping is important for And XXXV. 

Comparing to the CMDs of peer galaxies, And XXXV's apparent RHB or RC feature (at $\text{F814W}_0 \sim 24.6$\,mag) is more characteristic of M31 UFDs than of MW analogs. For example, MW UFD galaxies  CVn II, Hyd II, and Hydrus I with similar sizes and luminosities have generally bluer HBs. As opposed to similar MW systems, M31 dwarf galaxies tend to exhibit predominantly red HBs \citep[e.g.,][]{Martin_2017_And_redHB, Jennings_2023_UFD_HB_Ages}. Stellar population studies of M31 satellite galaxies indicate that such HB morphologies often correspond to extended SFHs, where star formation persisted until around $\sim$5$-$9\,Gyr ago \citep{Weisz2019, Savino2023_SFH}. This is the case even among the faintest M31 satellites, such as And XXVI, And XXII, and And XX, with $M_V \sim -$6.0, $-$6.4, and $-$6.4, respectively \citep{Martin2016}, which, although more luminous and spatially extended, exhibit similar RHB features to And XXXV. This age and metallicity spread both supports our classification of And XXXV as a galaxy and highlights that the stellar population differences between MW and M31 satellites persist well into the UFD regime at $M_V \sim -5$, where reionization was expected to play a very important role in shaping galactic SFHs \citep{Bovill2009,Brown2014,Bullock2017}. 

The discovery of Andromeda XXXV, the faintest and most compact satellite galaxy found in M31's halo to date, stresses the potential and importance of  uncovering faint and low-surface-brightness systems in the Local Volume and around MW-mass galaxies. Its seemingly complex stellar populations are important in this context, demonstrating that differences in SFH between MW and M31 (and more broadly, Local Group) UFDs persist to $M_V \sim -5$, deep in the regime where reionization was thought to dominate galactic SFHs. Given our shallow HST/WFC3 data, deeper follow-up imaging---e.g., with the James Webb Space Telescope (JWST) Near Infrared Camera, HST's Wide Field Camera 3, or the future Roman Space Telescope---could better resolve faint HB or RR-Lyrae stars (if they exist in sufficient numbers) for more precise distance determination and provide more detailed insights into And XXXV's complex stellar populations and SFH. Owing to the extreme faintness of And XXXV's stars, spectroscopic kinematic measurements to dynamically confirm the presence of dark matter (with $\sim$km\,s$^{-1}$ accuracy) are likely out of reach for the foreseeable future \citep{Simon2019}. However, long exposures with JWST NIRCam, HST narrowband imaging \citep[e.g.,][]{Fu_2022_NarrowBand_Met_EriII, Fu_2023_NarrowBand_Met_13_UFD_Cands}, or future northern hemisphere 30\,m class telescopes may yield accurate enough metallicities to confirm a metallicity spread. Looking ahead, upcoming facilities like the Rubin Observatory and Roman Space Telescope, with their unparalleled survey depth and coverage, combined with JWST and 30\,m class follow-up, promise to expand our census of UFDs, pushing detection limits to fainter magnitudes, smaller sizes, and a wider range of galactic environments. 

\begin{acknowledgments}
We thank the anonymous referee for comments and suggestions that improved the quality of this manuscript. This work was partly supported by HST grant GO-17158 provided by NASA through a grant from the Space Telescope Science Institute, which is operated by the Association of Universities for Research in Astronomy, Inc., under NASA contract NAS5-26555, the National Science Foundation through grant NSF-AST 2007065, and by the WFIRST Infrared Nearby Galaxies Survey (WINGS) collaboration through NASA grant NNG16PJ28C through subcontract from the University of Washington. A.M. gratefully acknowledges support by FONDECYT Regular grant 1212046 and by the ANID BASAL project FB210003, as well as funding from the Max Planck Society through a ``PartnerGroup” grant. This research has made use of NASA's Astrophysics Data System Bibliographic Services.
\end{acknowledgments}

\facility{HST}
\software{\code{DOLPHOT} \citep{dolphot2000, dolphot2016}, \code{Python 3.11} (\citealt{python}), \code{Matplotlib} (\citealt{matplotlib}), \code{Numpy} (\citealt{numpy}), \code{Scipy} (\citealt{scipy}), \code{scikit-learn} (\citealt{scikit-learn}), \code{scikit-image} (\citealt{scikit-image}), \code{Astropy} (\citealt{astropy1_2013, astropy2_2018AJ....156..123A, astropy3_2022ApJ...935..167A}), \code{dustmaps} (\citealt{dustmaps}), \code{emcee} \citep{Foreman-Mackey2013}, \code{Jupyter} Notebooks (\citealt{jupyter}).}

\bibliography{sample631}{}
\bibliographystyle{aasjournal}

\end{document}